\documentclass[%
 reprint,
 amsmath,amssymb,
 aps,
]{revtex4-2}

\usepackage{graphicx}
\usepackage{dcolumn}
\usepackage{bm}
\usepackage{cleveref}

\begin{document}

\iffalse
\newcommand{\REVISE}[1]{\color{red!80!black}{#1}\color{black}\xspace}
\newcommand{\CZY}[1]{\color{brown}{#1}\color{black}\xspace}
\newcommand{\TODO}[1]{\color{green}{#1}\color{black}\xspace}
\else
\newcommand{\REVISE}[1]{#1\xspace}
\newcommand{\CZY}[1]{#1\xspace}
\newcommand{\TODO}[1]{#1\xspace}
\fi

\Crefname{figure}{Fig.}{Figs.}
\Crefname{equation}{Eq.}{Eqs.}
\Crefname{algorithm}{Alg.}{Algs.}

\preprint{APS/123-QED}

\title{Neural Network-Based Frequency Optimization for Superconducting Quantum Chips}

\author{Bin-Han Lu\textsuperscript{1,2}} 
\author{Qing-Song Li\textsuperscript{1,2}}
\author{Peng Wang\textsuperscript{1,2}}
\author{Zhao-Yun Chen\textsuperscript{*3}}
\author{Yu-Chun Wu\textsuperscript{1,2,3}}
\author{Guo-Ping Guo\textsuperscript{1,2,3,4}}

\affiliation{\textsuperscript{1} Key Laboratory of Quantum Information Chinese Academy of Sciences, School of Physics, University of Science and Technology of China, Hefei, Anhui, 230026, P. R. China}
\affiliation{\textsuperscript{2} CAS Center For Excellence in Quantum Information and Quantum Physics, University of Science and Technology of China, Hefei, Anhui, 230026, P. R. China}
\affiliation{\textsuperscript{3} Institute of Artificial Intelligence, Hefei Comprehensive National Science Center, Hefei, Anhui, 230026, P. R. China}
\affiliation{\textsuperscript{4} Origin Quantum Co., Ltd. (Hefei)."}

\begin{abstract}
Optimizing the frequency configuration of qubits and quantum gates in superconducting quantum chips presents a complex NP-complete optimization challenge. This process is critical for enabling practical control while minimizing decoherence and suppressing significant crosstalk. In this paper, we propose a neural network-based frequency configuration approach. A trained neural network model estimates frequency configuration errors, and an intermediate optimization strategy identifies optimal configurations within localized windows of the chip. The effectiveness of our method is validated through randomized benchmarking and cross-entropy benchmarking. Furthermore, we found that for variational quantum eigensolvers, optimizing the frequency configurations for a crosstalk-aware hardware-efficient ansatz leads to improved energy calculations.
\end{abstract}

\maketitle



\begin{figure*}[htbp]
\centering
\begin{minipage}[t]{0.22\textwidth}
\centering
\includegraphics[width=\textwidth]{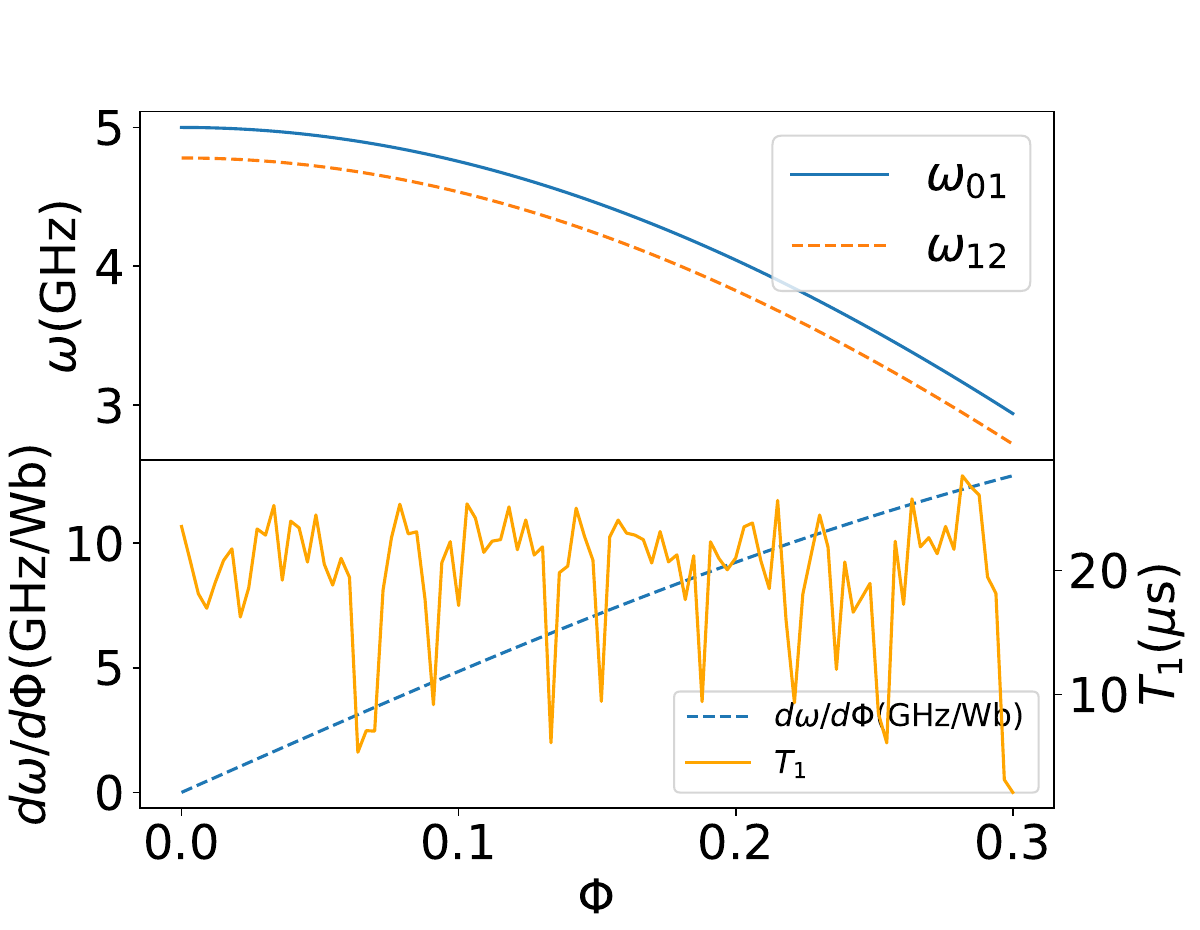}
(a)
\end{minipage}
\hspace{0.25cm} 
\begin{minipage}[t]{0.15\textwidth}
\centering
\includegraphics[width=\textwidth]{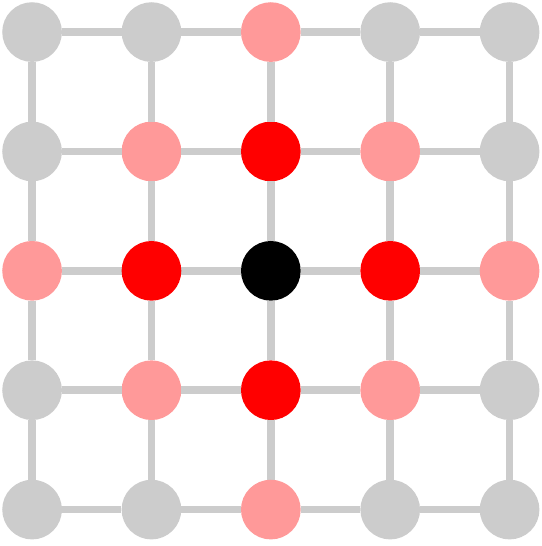}
(b)
\end{minipage}
\hspace{0.5cm} 
\begin{minipage}[t]{0.18\textwidth}
\centering
\includegraphics[width=\textwidth]{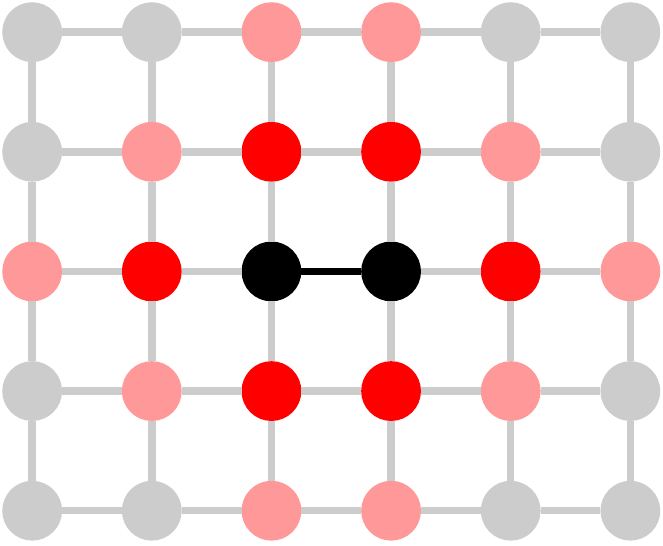}
(c)
\end{minipage}
\hspace{0.5cm} 
\begin{minipage}[t]{0.3\textwidth}
\centering
\includegraphics[width=\textwidth]{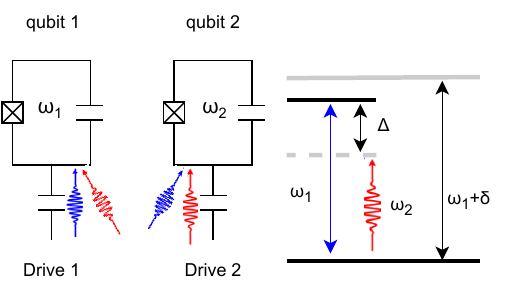}
(d)
\end{minipage}
\caption{(a) The frequency spectrum of tunable qubits, along with the variations in 
$T_1$  and $\frac{d\omega}{d\phi}$, as the external control magnetic flux changes.
(b-c) Coupling diagrams of quantum chips, where the dots represent qubits and the lines denote couplers. The black qubits (gate qubits) are shown to have stray coupling with all the red qubits.
(d) Illustration of microwave crosstalk, where the drive applied to the target qubit $q_1$ induces crosstalk with $q_2$. When the drive is not resonant with $q_2$, it causes a shift in the energy levels of $q_2$,  and this effect becomes more pronounced as the frequencies of $q_1$ and $q_2$ approach resonance.}\label{fig freq range}
\end{figure*}

To achieve fault-tolerant quantum computing \cite{horsman2012surface,chamberland2020topological,jochym2014using,paetznick2013universal}, superconducting quantum processors must scale beyond the limitations of noisy intermediate-scale quantum (NISQ) chips \cite{boixo2018characterizing}. However, this progression faces two significant challenges: unreliable hardware manufacturing and imperfections in control systems, which contribute to various errors. Decoherence errors, including qubit relaxation and dephasing, are one major issue \cite{ithier2005decoherence}. Another critical problem is crosstalk errors, which occur when excessive residual coupling between qubits causes parallel operations to interfere with each other \cite{zhao2022quantum,krinner2020benchmarking,seo2021mitigation}. Additionally, microwave pulses intended for single-qubit gates have unwanted effects on neighboring qubits, leading to microwave crosstalk errors \cite{zhao2022spurious}. Addressing these challenges is essential for enabling scalable fault-tolerant quantum computing.

The frequency-tunable architecture enables two-qubit gates by bringing neighboring qubits directly into resonance \cite{yan2018tunable,xu2020high,sung2021realization}, resulting in shorter gate execution times than fixed-frequency qubits. However, errors in this architecture are susceptible to the frequency configuration. Different configurations lead to variations in qubit dephasing and relaxation times. Notably, when qubit frequencies align with Two-Level System (TLS) defects \cite{tian2011cavity} or flux-sensitive points \cite{krantz2019quantum}, decoherence times are significantly reduced. Furthermore, crosstalk arises when qubit frequencies unintentionally come into resonance \cite{lu2024camelphysicallyinspiredcrosstalkaware}. As a result, optimizing qubit frequency configurations is critical to mitigating errors in frequency-tunable architectures \cite{brink2018device,klimov2020optimizing,klimov2020snake}.

Addressing frequency configuration requires developing a model that maps the frequency of each quantum gate to its associated error values. This model must account for decoherence noise at different qubit frequencies and identify whether qubit frequencies lie within closely resonant regions. Understanding how various errors interact and collectively impact qubits adds to the complexity.
Building upon this model, solving an optimization problem becomes necessary to determine a set of frequency configurations that minimize errors for each quantum gate. However, as the number of qubits increases, the scale of the problem grows exponentially, making it a highly constrained and computationally intensive challenge.

Existing research tackles these challenges using compensation pulses to mitigate XY control signal crosstalk \cite{krinner2022realizing} and two-tone flux modulation to suppress decoherence \cite{valery2022dynamical}. However, these methods inevitably increase the complexity of the control system. Given the strong dependence of errors on frequency, a more direct approach involves identifying frequency configurations that circumvent hardware imperfections, thereby simplifying hardware control.

While some frequency configuration strategies lack a thorough investigation of the underlying physical systems \cite{versluis2017scalable, Ding2020SystematicCM}, Google’s frequency configuration scheme \cite{klimov2024optimizing} offers a more advanced solution by incorporating a deeper understanding of the physical principles. This approach employs an error model that linearly combines all sources of error, demonstrating a more comprehensive grasp of the system. Nonetheless, it has certain limitations. First, the impact of different error sources on qubits may not be accurately captured by linear combinations. Second, error sources such as gate distortion, microwave crosstalk, and other unknown global errors remain challenging to model precisely \cite{klimov2020optimizing}, which limits the generalization and scalability of this method. 
In \cite{Ai2024amx}, Ai et al. proposed using a multilayer perceptron to predict gate errors and a graph neural network for frequency optimization. After training, the multilayer perceptron is used to predict errors on other chips. However, due to fabrication inconsistencies, the error sources of each quantum chip vary, and the trained neural network may lack generalizability.
In \cite{morvan2022optimizing, zhang2024efficient}, Morvan et al. and Zhang et al. formulated the frequency configuration problem as an integer programming problem with linear constraints. However, the physical mechanisms of qubit noise cannot be adequately captured by a simple set of linear constraints. Moreover, in addition to the mutual interactions between qubits, each qubit also exhibits its own complex noise mechanisms.

In this paper, we propose a neural network-based frequency configuration scheme to address these challenges. Unlike methods that rely on linear combinations of errors, neural networks can effectively learn and model the nonlinear interactions among error mechanisms. Additionally, as model-free tools, neural networks adapt to complex and dynamic environments, enabling them to better capture difficult-to-quantify error sources such as microwave crosstalk and gate distortions. Moreover, our approach eliminates the need to collect calibration data, such as decoherence times, for building the error model, significantly simplifying the data collection process.

In our frequency allocation process, we begin by randomly initializing a frequency configuration and using the neural network to predict gate errors across the chip. The window with the highest average error is identified and optimized locally. This iterative process is repeated until the overall average gate errors are minimized. In contrast, Google’s scheme starts without predefined frequencies, progressively configuring windows adjacent to the previous step until all gates are assigned. However, their approach does not allow modifications to previously set frequencies, potentially leading to higher overall errors. Our method, by enabling iterative revisions to prior configurations, achieves lower global average errors and greater optimization flexibility.

Finally, after completing the configuration, we conducted single-qubit randomized benchmarking (RB) \cite{knill2008randomized} and two-qubit cross-entropy benchmarking (XEB) \cite{arute2019quantum} on the chip. The results significantly reduced gate errors, confirming that our frequency configuration effectively identifies and selects low-error-rate frequencies.
Furthermore, we developed optimized frequency configurations for crosstalk-aware hardware-efficient ansatz (HEA) \cite{kandala2017hardware,leone2022practical,wang2023eha} in variational quantum eigensolvers (VQE) \cite{du2021accelerating,wang2019accelerated,peruzzo2014variational}. Experimental results show that, under the same HEA, optimized frequency configurations achieve lower energy values compared to random configurations, highlighting the importance of frequency configurations in minimizing errors and enhancing computation fidelity.

\begin{figure*}[htbp]
    \centering
    \includegraphics[width=0.8\textwidth]{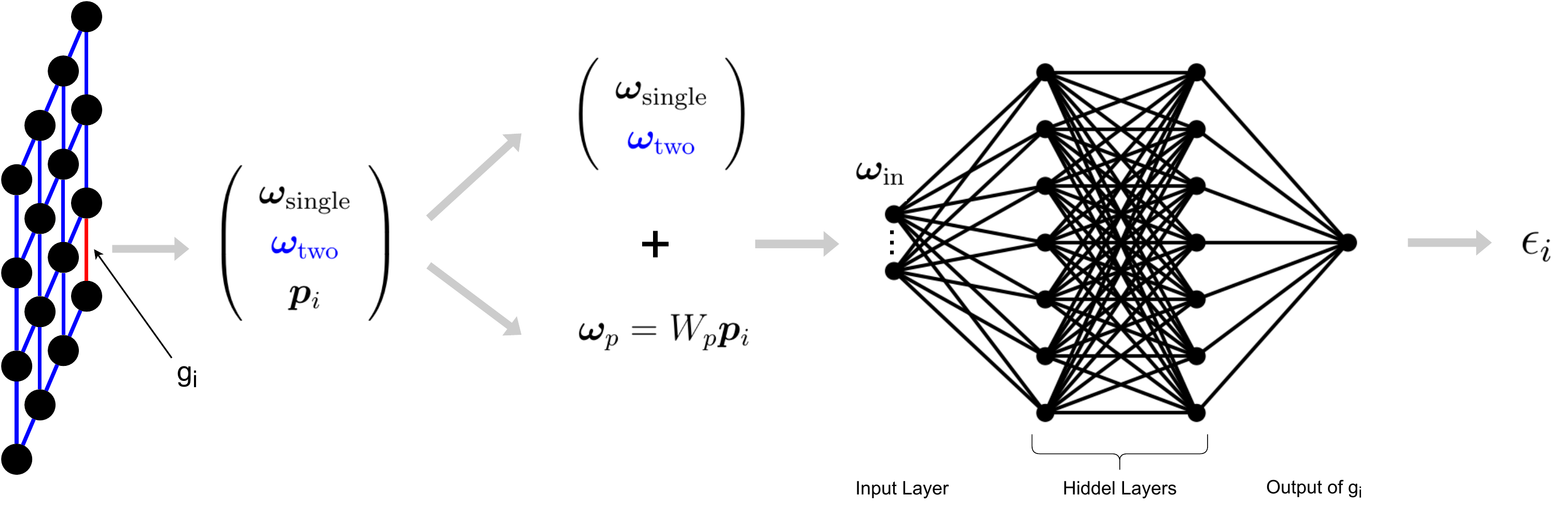}
    \caption{Given a chip and a set of frequency configuration $(\bm{\omega}_{\text{single}},\bm{\omega}_{\text{two}})$. The input vector is defined as $(\bm{\omega}_{\text{single}},\bm{\omega}_{\text{two},}\bm{p}_i)$. The vector $\bm{p}_i$  undergoes a linear transformation $W_p$, transforming it into a dense vector $\bm{\omega}_p$ incorporating positional information. This transformed vector is then added to $(\bm{\omega}_{\text{single}},\bm{\omega}_{\text{two}})$, resulting in a combined vector $\bm{\omega}_{\text{in}}$ that contains both configuration and positional information. $\bm{\omega}_{\text{in}}$ is then fed into a multilayer perceptron, and the final output $\epsilon_i$ is the predicted error of $g_i$ under configuration $(\bm{\omega}_{\text{single}},\bm{\omega}_{\text{two}})$.}\label{fig nnflow}
\end{figure*}

\begin{figure*}[htbp]
\centering
\begin{minipage}[t]{0.3\textwidth}
\centering
\includegraphics[width=\textwidth]{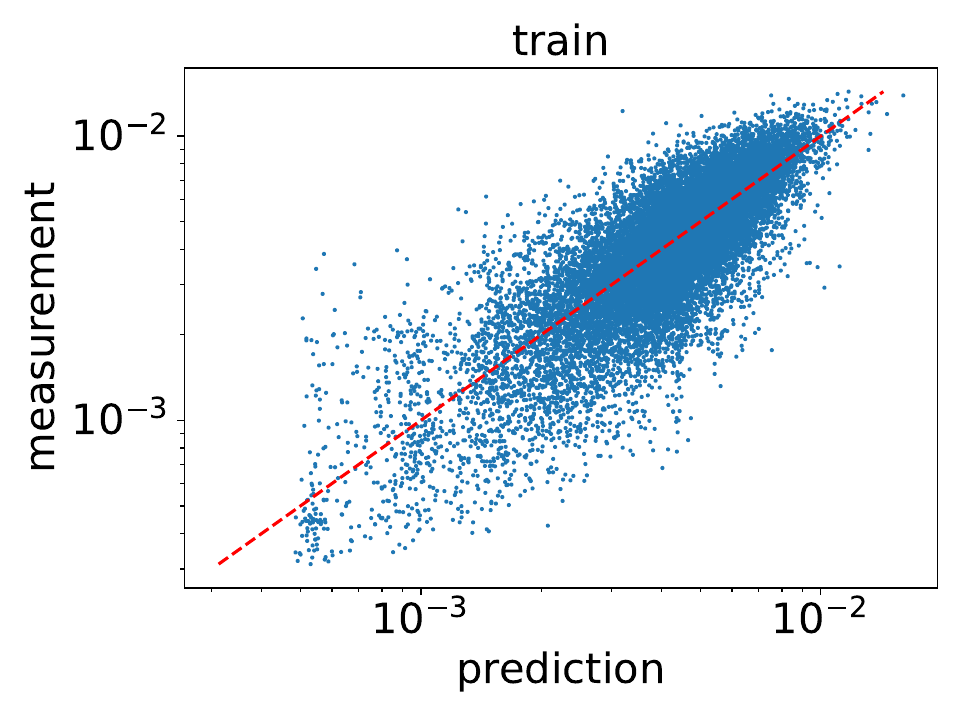}
(a)
\end{minipage}
\begin{minipage}[t]{0.3\textwidth}
\centering
\includegraphics[width=\textwidth]{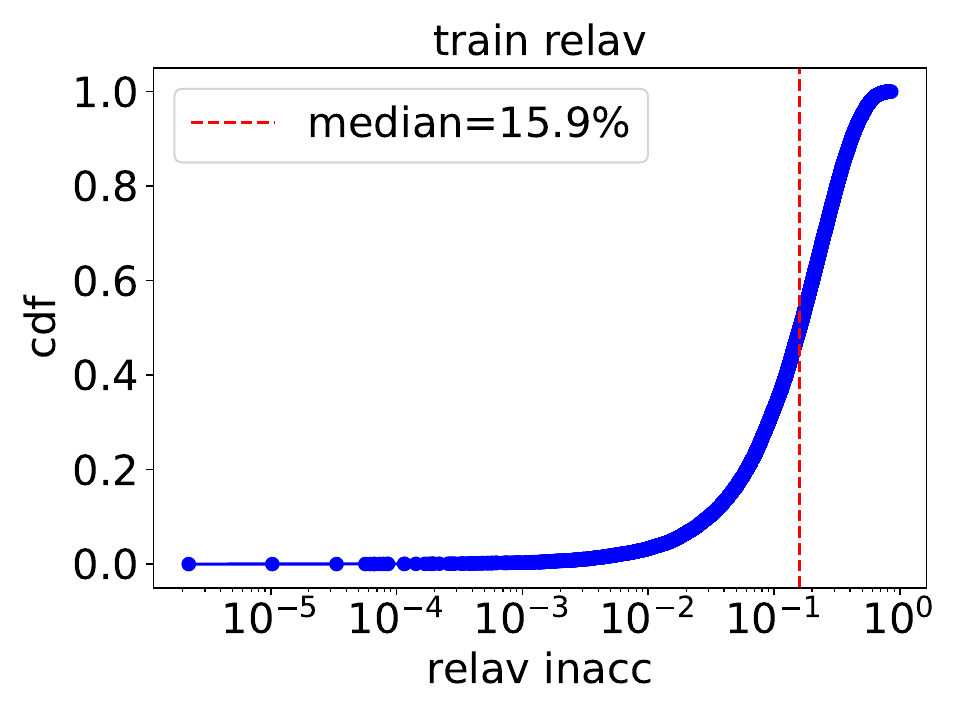}
(b)
\end{minipage}
\begin{minipage}[t]{0.3\textwidth}
\centering
\includegraphics[width=\textwidth]{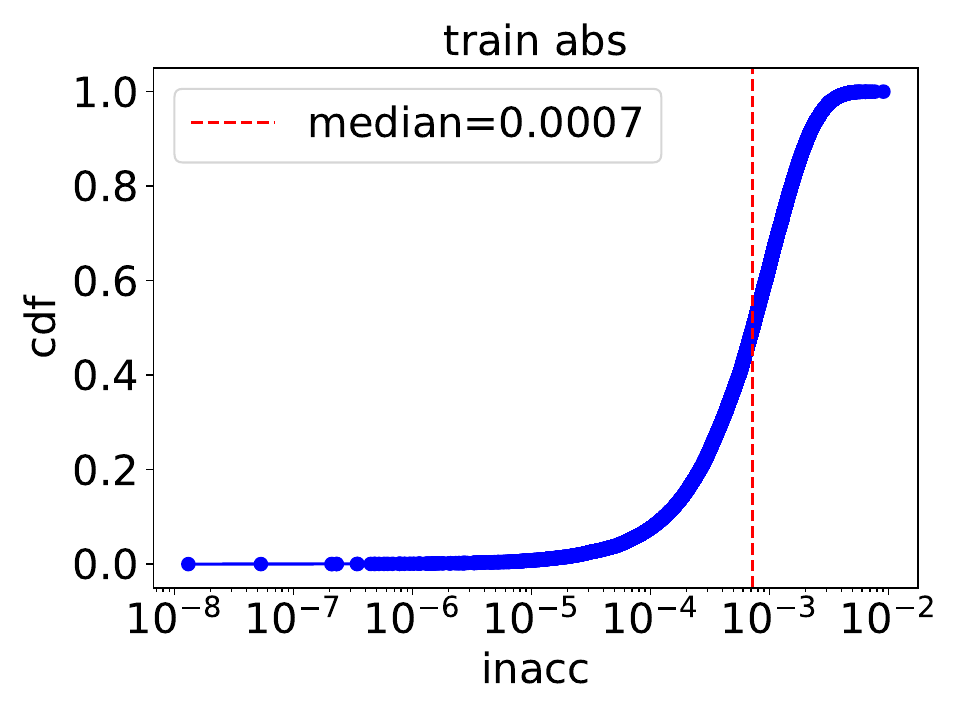}
(c)
\end{minipage}
\begin{minipage}[t]{0.3\textwidth}
\centering
\includegraphics[width=\textwidth]{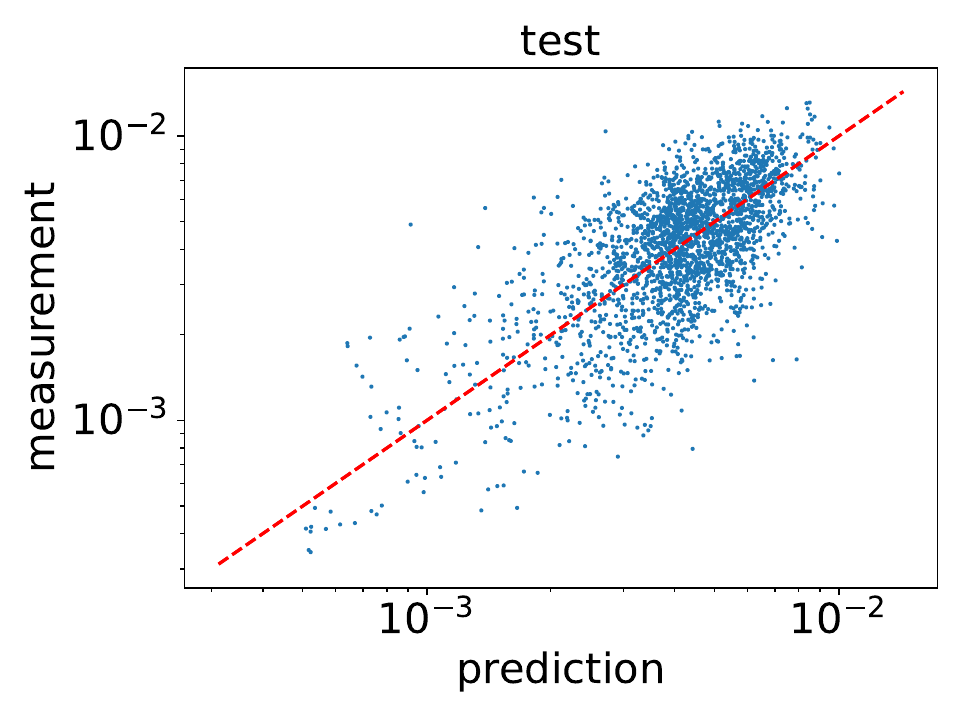}
(d)
\end{minipage}
\begin{minipage}[t]{0.3\textwidth}
\centering
\includegraphics[width=\textwidth]{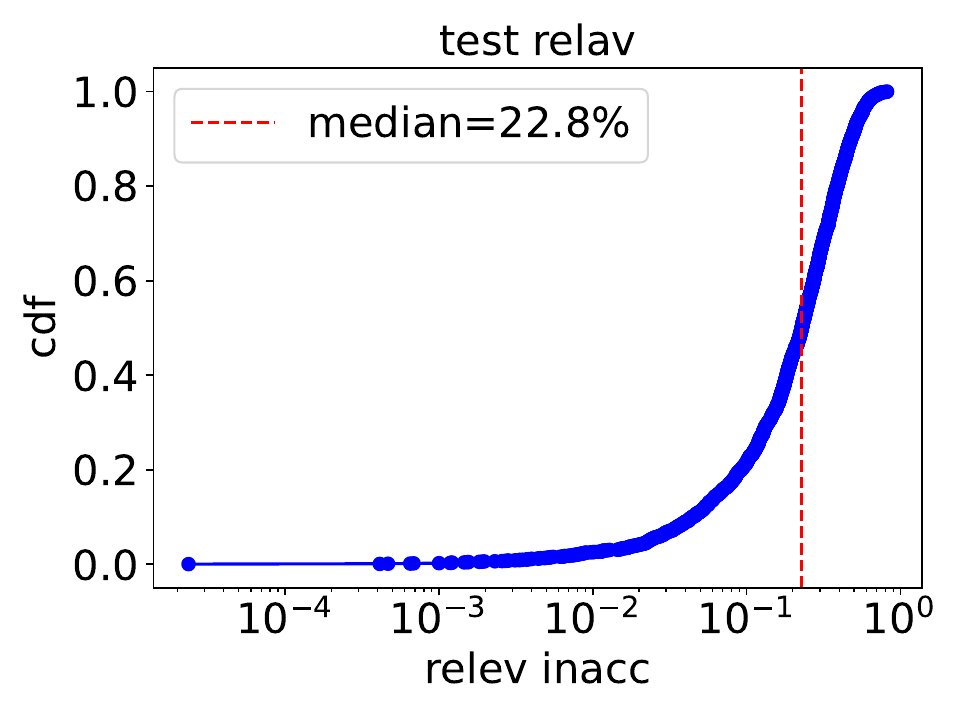}
(e)
\end{minipage}
\begin{minipage}[t]{0.3\textwidth}
\centering
\includegraphics[width=\textwidth]{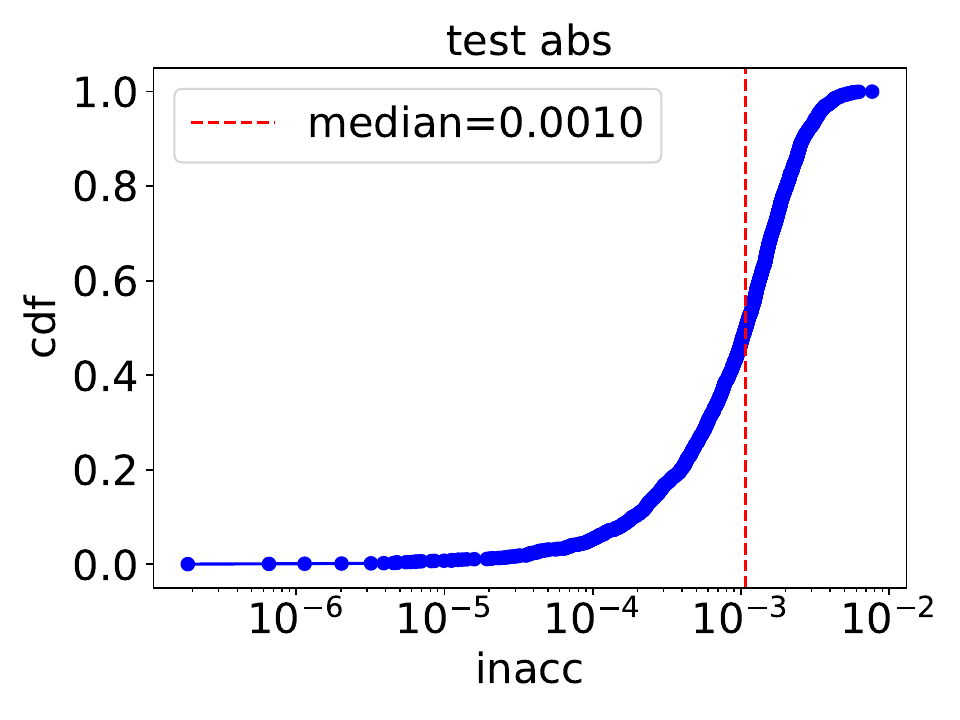}
(f)
\end{minipage}
\caption{
(a) Scatter plot of the training set, showing the measured error data (vertical axis) and the corresponding predicted error data (horizontal axis) for each gate under a given configuration. Ideally, all points should cluster around the red line, where $\lvert\text{prediction} - \text{measure}\rvert = 0$. Similarly, (d) presents the scatter plot for the test set, where points in both the training and test sets are distributed around the red line after training.
(b) Cumulative frequency distribution plot of relative errors, $|\text{prediction} - \text{measure}|/\text{measure}$, in the training set, with a median of $15.9\%$.
(e) Cumulative distribution plot of relative errors in the test set, with a median of $22.8\%$.
(c) Cumulative frequency distribution plot of absolute errors, $|\text{prediction} - \text{measure}|$, in the training set, with a median of $7\times 10^{-4}$.
(f) Cumulative distribution plot of absolute errors in the test set, with a median of $1\times 10^{-3}$.
}\label{fig train test}
\end{figure*}

\textbf{Error Mechanisms}
On a quantum chip, the frequencies of single-qubit and two-qubit gates directly impact the final computational results. 
The first type of error is relaxation error ($T_1$ error) \cite{ithier2005decoherence}. As the qubit frequency shifts, its $T_1$ relaxation time also changes. 
The $T_1$ relaxation time sharply decreases at certain Two-Level System (TLS) defect points \cite{tian2011cavity}, causing the qubit to easily transition from the higher energy state to the lower energy state.
The second type is dephasing error ($T_2$ error) \cite{krantz2019quantum}. In a tunable-frequency architecture, a qubit's dephasing time 
$T_2$ depends on the first derivative of the frequency to the magnetic flux, given by $\frac{1}{T_2}\propto\frac{d\omega}{d\phi}$  (see \Cref{fig freq range}(a)). Therefore, frequency settings should be chosen such that both $T_1$ and $T_2$ are sufficiently long to minimize errors (see \Cref{fig freq range}(a)). The third type of error occurs during two-qubit gate execution when the qubit frequency shifts from the idle frequency $\omega_\text{off}$ to the interaction frequency $\omega_\text{on}$. An excessive shift $\lvert\omega_\text{off}-\omega_\text{on}\rvert$ can cause gate distortion errors. The fourth type is stray coupling crosstalk \cite{klimov2020optimizing}. 
As shown in \Cref{fig freq range}(b-c), if the gate-qubit frequency approaches resonance with neighboring or next-neighboring qubits, unintended coupling occurs, which can reduce gate fidelity.
Finally, microwave crosstalk (see \Cref{fig freq range}(d)) arises when a drive signal intended for qubit $q_i$ influences a nearby qubit $q_j$. If $q_i$ and $q_j$ are near resonance, $q_j$ is significantly affected. Unlike stray coupling, microwave crosstalk has a broader reach and can impact even non-neighboring qubits.

\begin{figure*}[htbp]
\centering
\begin{minipage}[t]{0.06\textwidth}
\centering
\includegraphics[width=\textwidth]{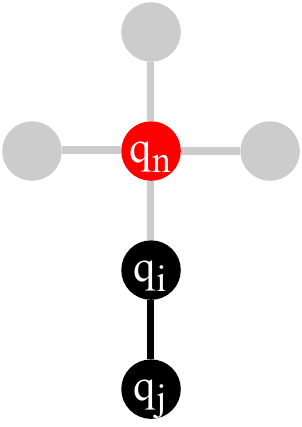}
(a)
\end{minipage}
\hspace{0.3cm} 
\begin{minipage}[t]{0.18\textwidth}
\centering
\includegraphics[width=\textwidth]{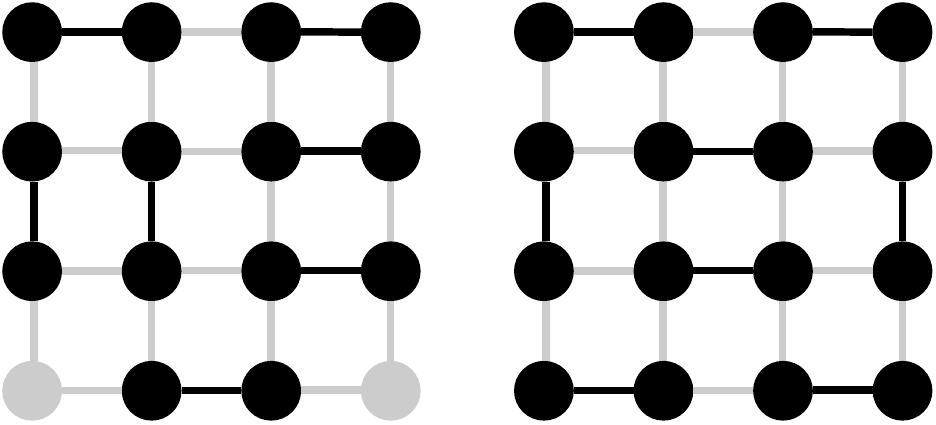}
(b)
\end{minipage}
\hspace{0.3cm} 
\begin{minipage}[t]{0.38\textwidth}
\centering
\includegraphics[width=\textwidth]{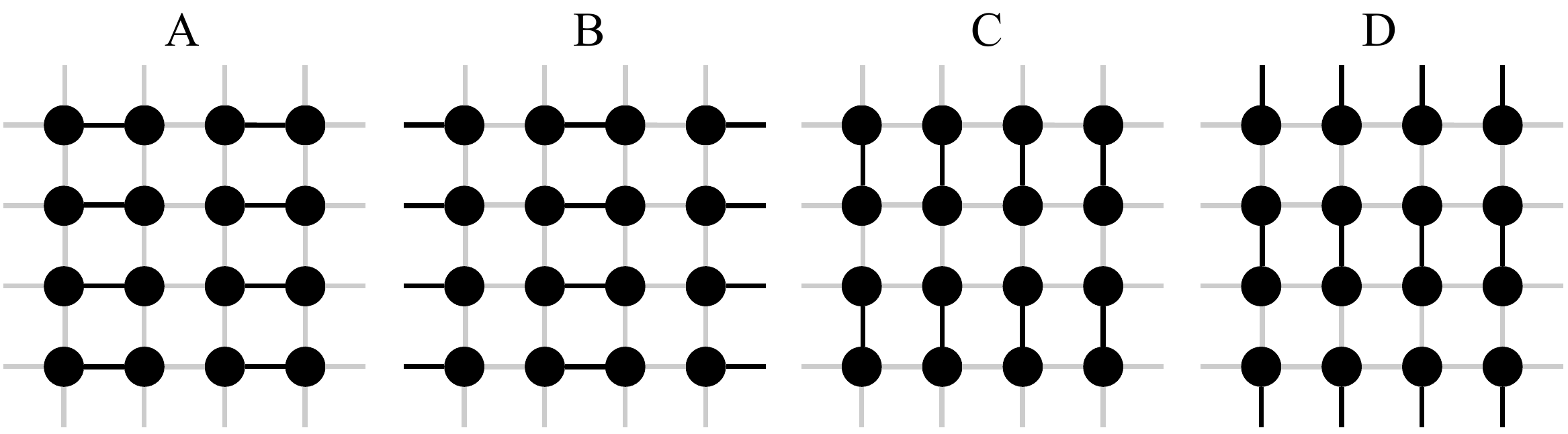}
(c)
\end{minipage}
\hspace{0.3cm} 
\begin{minipage}[t]{0.2\textwidth}
\centering
\includegraphics[width=\textwidth]{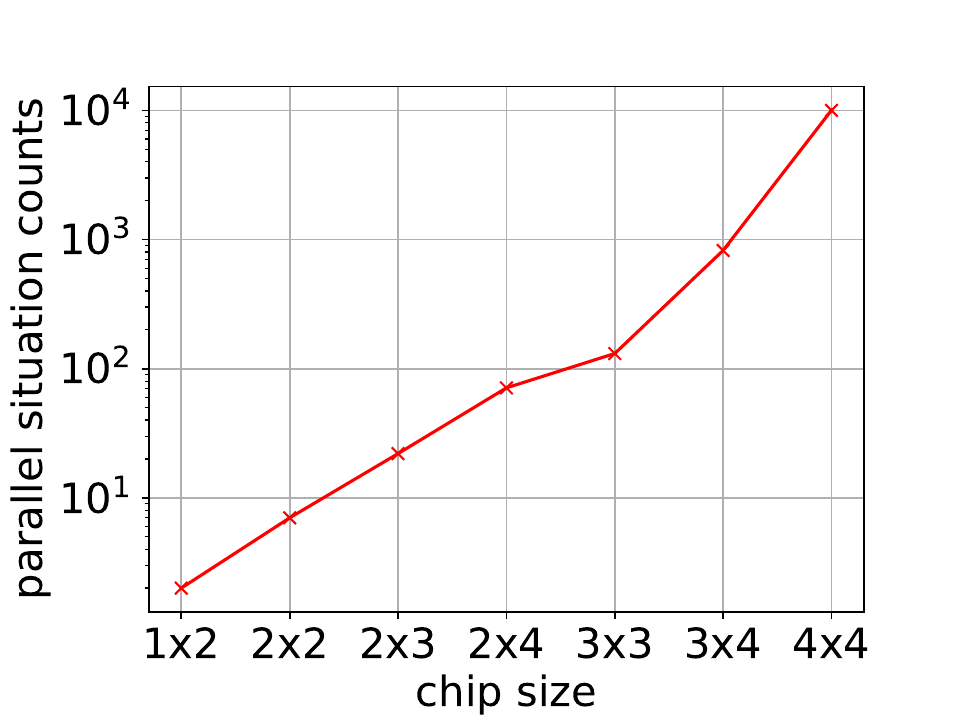}
(d)
\end{minipage}
\caption{(a) Qubits $q_i$ and $q_j$  perform a two-qubit gate, with $q_n$ as the spectator. In a periodic grid, $q_n$ can either execute a two-qubit gate with one of its three neighboring qubits (represented by gray dots) or a single-qubit gate, resulting in four possible frequency configurations. Considering all four scenarios introduces excessive constraints, expanding the exclusion zones.
(b) The black edges and nodes represent the qubits and couplers involved in quantum gates within a specific time slice. Four example configurations are shown.
(c) The coupler activation pattern determines which qubits can execute two-qubit gates simultaneously.
(d) The number of parallel two-qubit gate scenarios increases with the scale of the $M\times N$ chip.}
\label{fig parallel sit}
\end{figure*}

\textbf{Neural Network Error Estimator}
We use a neural network to predict gate error under a given frequency configuration \cite{wu2018development,dongare2012introduction}. The network (see \Cref{fig nnflow}) consists of a position embedding layer \cite{huang2020improve}, an input layer, hidden layers, and an output layer. Overall, it is a multilayer perceptron \cite{popescu2009multilayer,gardner1998artificial}.

The network uses the frequencies of all single-qubit and two-qubit gates as inputs and predicts the errors for a specific single-qubit or two-qubit gate under a given configuration. Based on the previously discussed error mechanisms, the error of a single-qubit gate is influenced by its own frequency, the frequencies of its neighboring and next-neighboring qubits, as well as the frequencies of qubits inducing microwave crosstalk.
For a two-qubit gate, its error is affected by the idle and interaction frequencies of the gate qubits, the frequencies of neighboring spectator qubits, and the frequencies of qubits causing microwave crosstalk with the gate qubits.
To determine which qubit frequencies affect the target quantum gate, the neural network analyzes the input data and dynamically refines the connections between its neurons to optimize the learning process. As a result, the input vector must comprehensively include the single-qubit and two-qubit frequencies of all qubits on the chip.
\begin{equation}
    \bm{\omega}_{\text{in}}=(\bm{\omega}_{\text{single}},\bm{\omega}_{\text{two}}).
\end{equation}
Here, $\bm{\omega}_{\text{single}}$ and $\bm{\omega}_{\text{two}}$ represent the frequencies of single-qubit and two-qubit gates, respectively. For preprocessing the input data, let  $\omega_{\min}$ and $\omega_{\max}$  denote the minimum and maximum frequencies across all single-qubit and two-qubit gates. Normalize the input data by mapping the frequency range to the interval $(0, 1)$.

To enable the neural network to predict the error for a specific quantum gate, the input vector must include information about the gate's position on the chip. This is achieved using position embedding \cite{huang2020improve}. For clarity of explanation, we consider an example chip with an $M\times N$ qubit array and $2MN-M-N$ couplers, resulting in an input vector of dimension $3MN-M-N$. For position embedding, we define a one-hot sparse vector $\bm{p}_i$ with the same dimension as the number of gates ($3MN-M-N$ in this example). To predict the error $\epsilon_i$ for the $i$-th gate, $\bm{p}_i$ is set such that  $(\bm{p}_i)_j=\delta_{ij}$.
$\bm{p}_i$ is transformed via a trainable linear layer $W_p$, yielding a dense vector $\bm{\omega}_p=W_p\bm{p}_i$ with dimensions matching the input vector $\bm{\omega}_{\text{in}}$. The final input to the neural network is constructed by combining $\bm{\omega}_p$ and $\bm{\omega}_{\text{in}}$  through addition: $\bm{\omega}=\bm{\omega}_{\text{in}}+\bm{\omega}_p$. This combined vector encodes both the frequency configuration and positional information, enabling the network to output the error $\epsilon_i$ for the target gate. The trainable position embedding layer is effective for this network as it processes fixed-dimension inputs without requiring extrapolation.

\Cref{fig train test}(a-c) shows the training results. In \Cref{fig train test}(a), the scatter plot of predicted errors versus measured errors is shown, with ideal data points aligning closely along the diagonal. \Cref{fig train test}(b) and \Cref{fig train test}(c) present cumulative frequency distribution plots of the relative and absolute errors between the predicted and measured values.
\Cref{fig train test}(d-f) illustrates the test results. As seen in \Cref{fig train test}(e) and \Cref{fig train test}(f), the median relative error and absolute error in the test set are $22.8\%$ and $1\times 10^{-3}$, respectively. These values closely match those in the training set and fall within the same order of magnitude as the results from Google's model. Notably, our test set includes only 500 configurations, significantly fewer than Google's 6500 configurations.
\begin{figure*}[htbp]
\centering
\begin{minipage}[t]{0.18\textwidth}
\centering
\includegraphics[width=\textwidth]{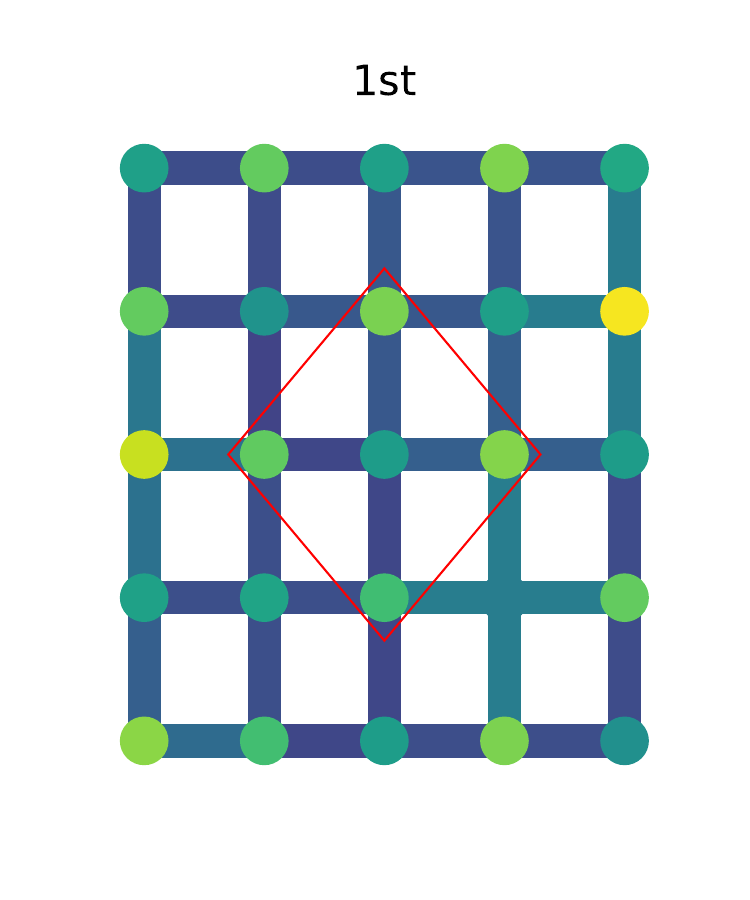}
\end{minipage}
\hspace{-0.7cm}
\begin{minipage}[t]{0.18\textwidth}
\centering
\includegraphics[width=\textwidth]{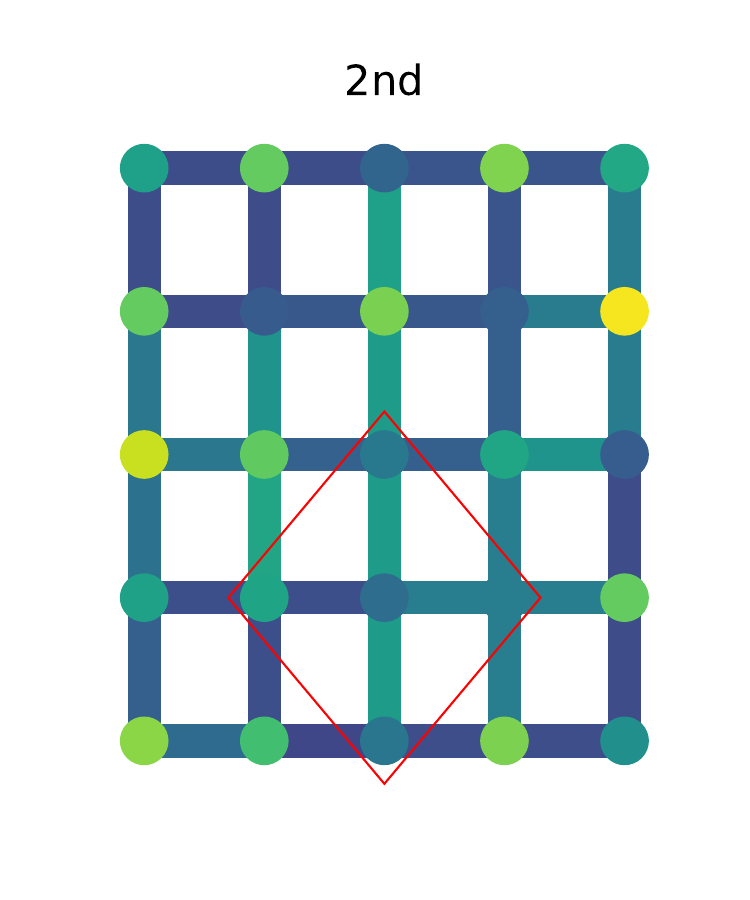}
(a)
\end{minipage}
\hspace{-0.7cm}
\begin{minipage}[t]{0.18\textwidth}
\centering
\includegraphics[width=\textwidth]{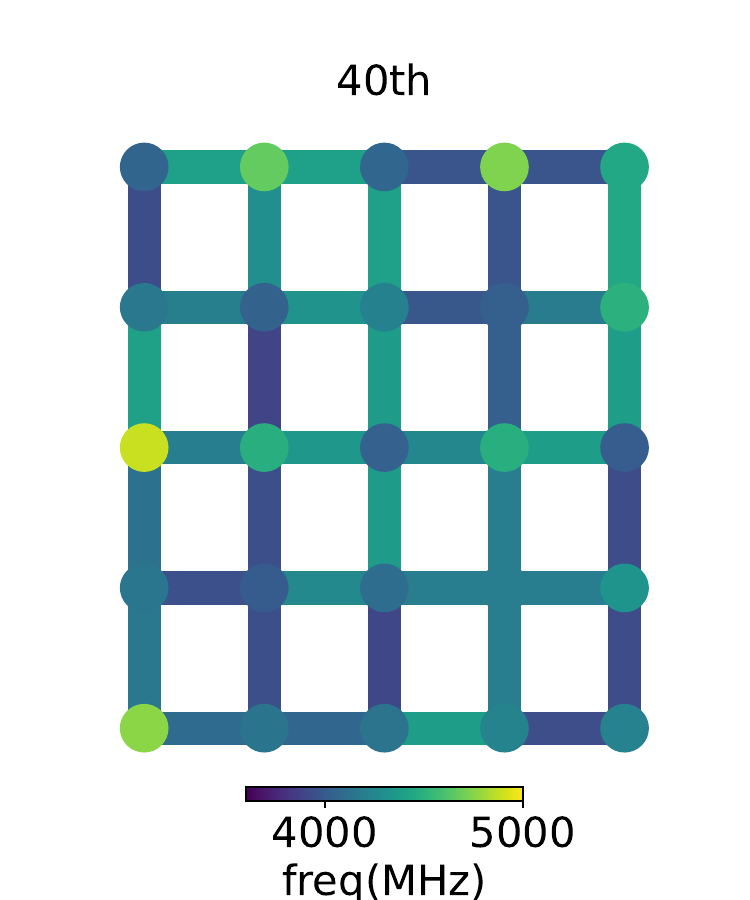}
\end{minipage}
\hspace{-0.7cm}
\begin{minipage}[t]{0.18\textwidth}
\centering
\includegraphics[width=\textwidth]{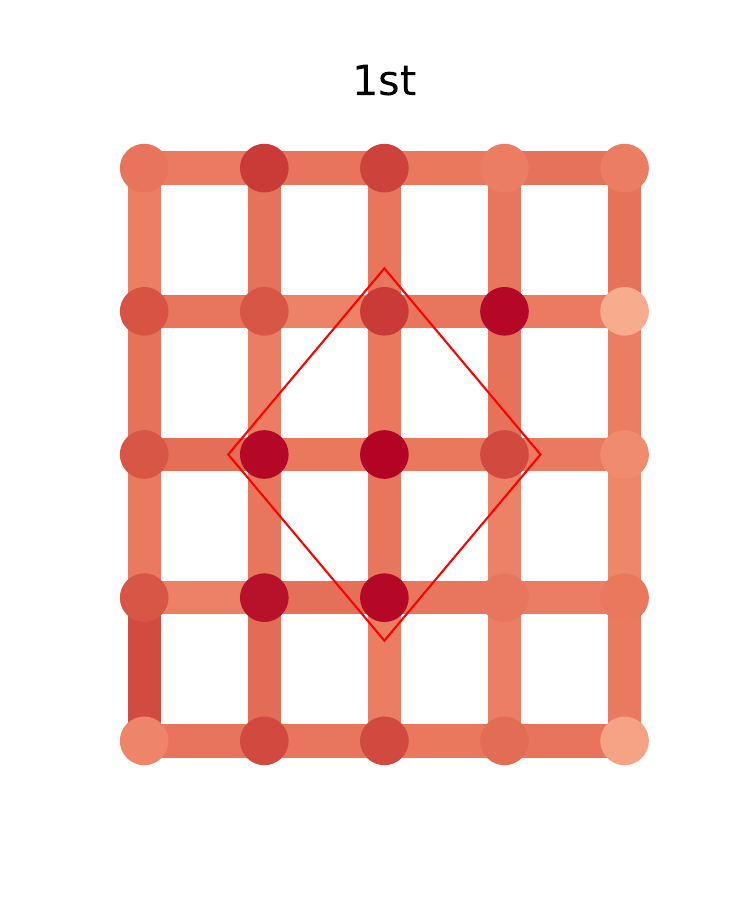}
\end{minipage}
\hspace{-0.7cm}
\begin{minipage}[t]{0.18\textwidth}
\centering
\includegraphics[width=\textwidth]{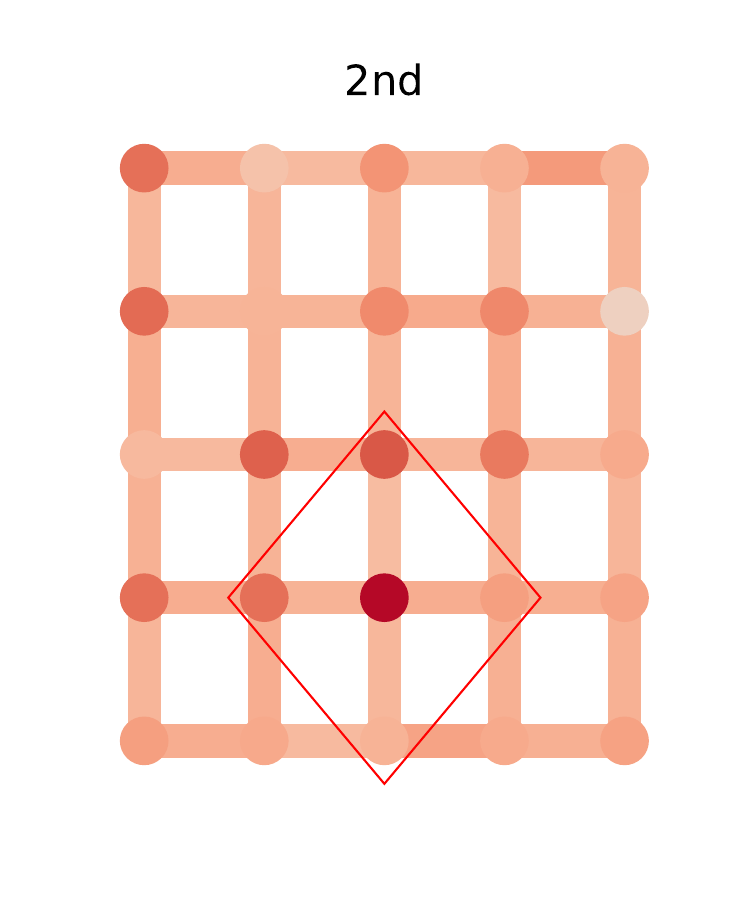}
(b)
\end{minipage}
\hspace{-0.7cm}
\begin{minipage}[t]{0.18\textwidth}
\centering
\includegraphics[width=\textwidth]{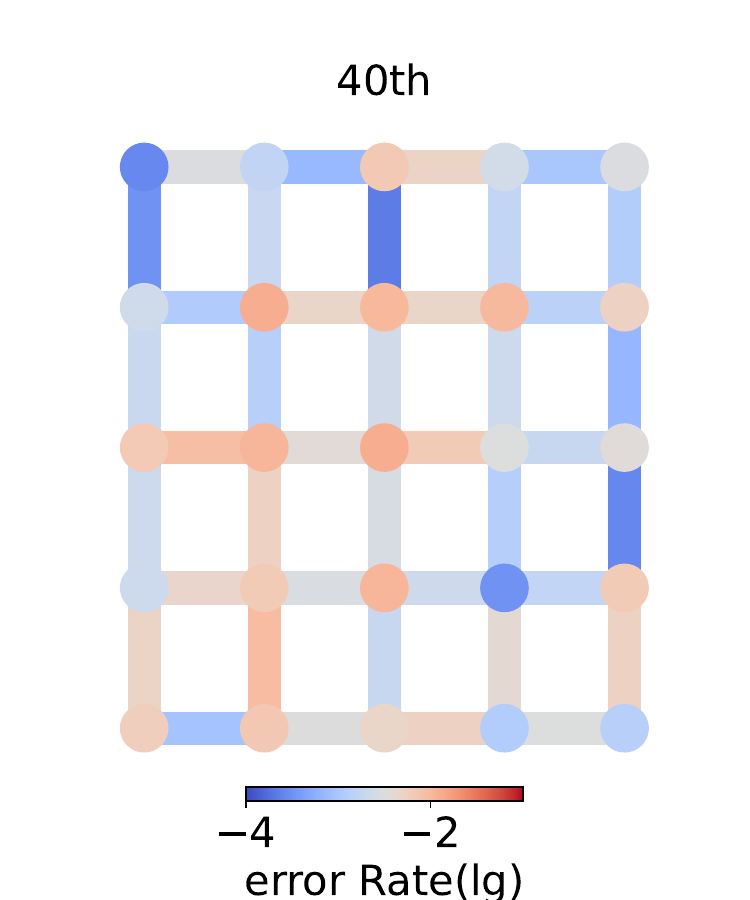}
\end{minipage}
\begin{minipage}[t]{0.25\textwidth}
\centering
\includegraphics[width=\textwidth]{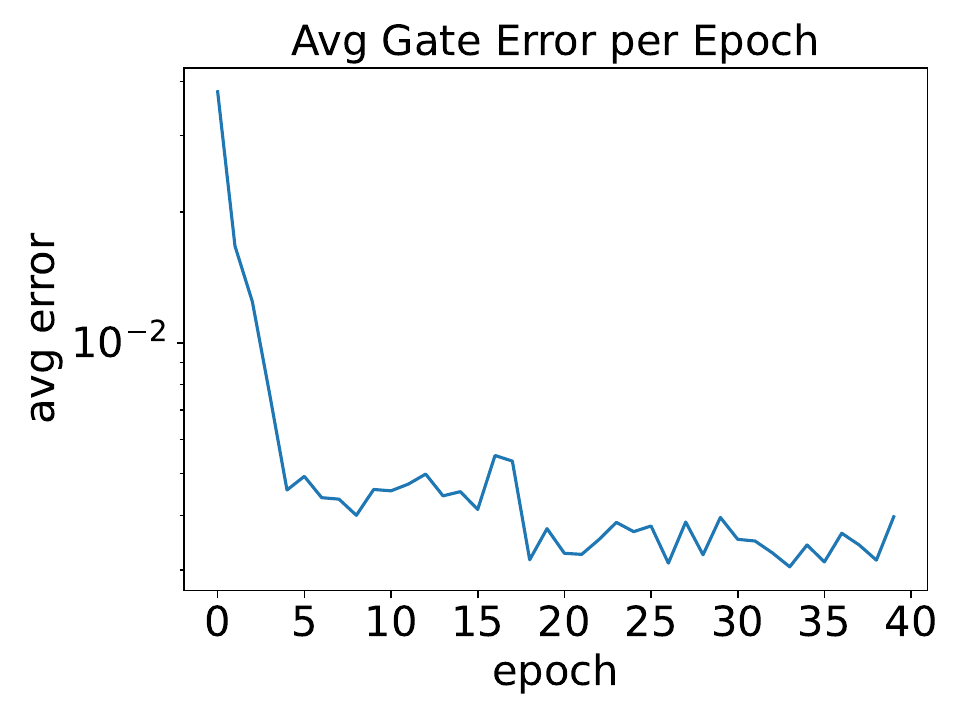}
(c)
\end{minipage}
\begin{minipage}[t]{0.25\textwidth}
\centering
\includegraphics[width=\textwidth]{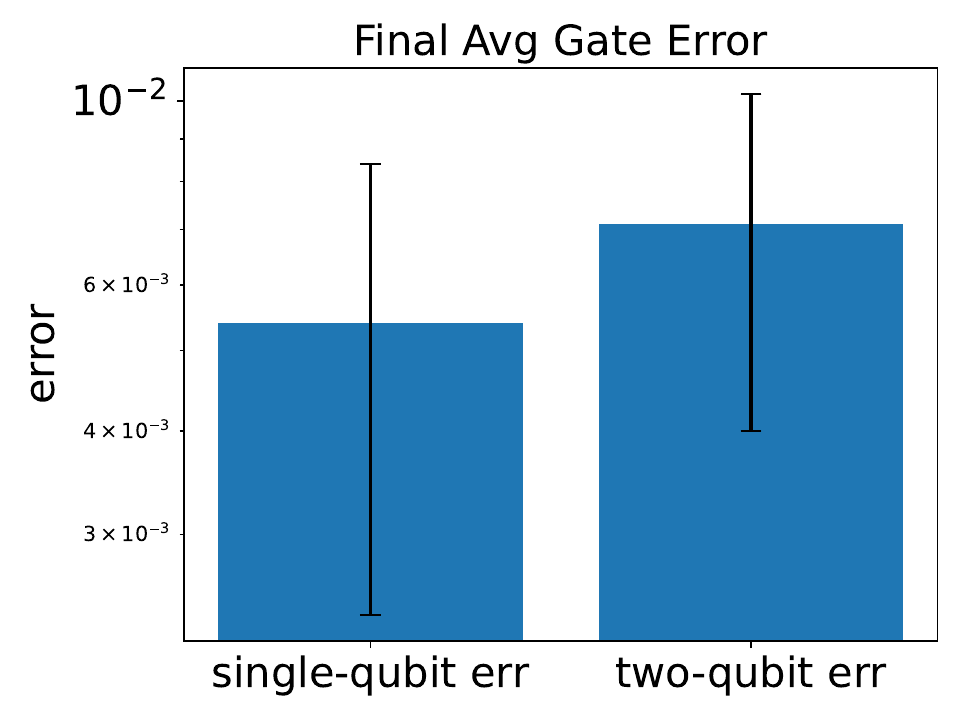}
(d)
\end{minipage}
\begin{minipage}[t]{0.25\textwidth}
\centering
\includegraphics[width=\textwidth]{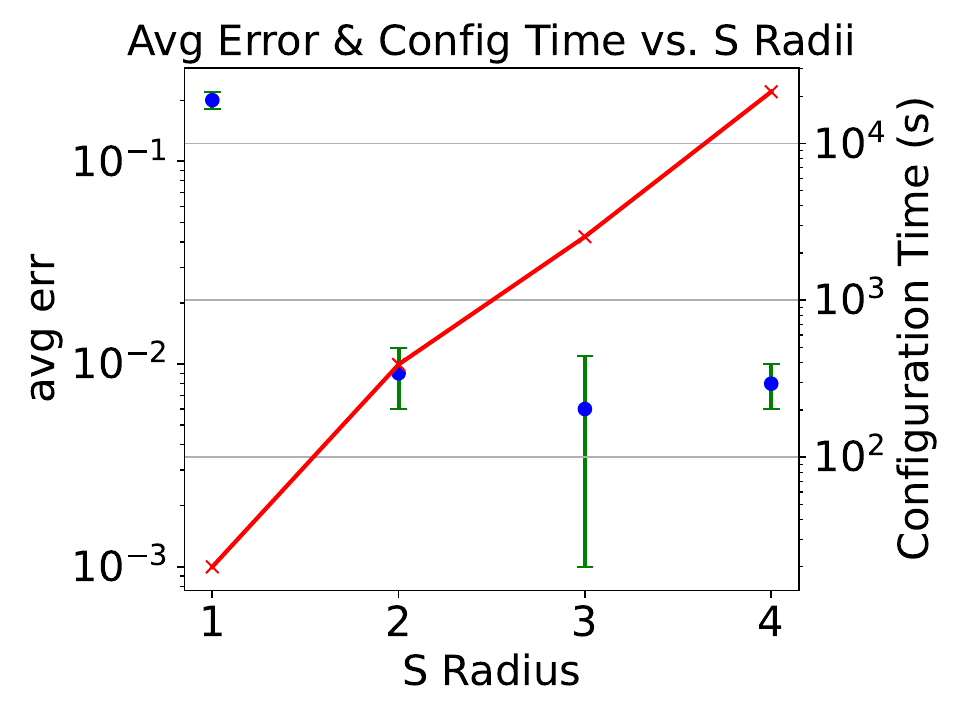}
(e)
\end{minipage}
\caption{
(a-b) From left to right, the frequencies and errors are shown for the 1st, 2nd, and 40th(last) iterations, respectively. 
In each instance, the region $S$ with the highest average error is selected (indicated by a red diamond).
(c) The overall average error decreases progressively with the number of iterations.
(d) The average errors of single-qubit and two-qubit gates are displayed.
(e) The average gate error decreases as the radius of $S$ increases.
}\label{fig freq err}
\end{figure*}

\textbf{Frequency Configuration Strategy}
In a periodic grid structure (see \Cref{fig parallel sit}(a)), when qubits $q_i$ and $q_j$ execute a two-qubit gate, their neighboring qubit $q_n$ may engage in a two-qubit gate with up to three other qubits or a single-qubit gate, requiring up to four distinct frequency settings. Ideally, an optimized frequency configuration should avoid any crosstalk between parallel quantum gates. However, for an $M \times N$ chip (see \Cref{fig parallel sit}(b)), the number of couplers is $2MN - M - N$, with each two-qubit gate on a coupler having two possible states: executed or not executed. Consequently, the number of potential parallel two-qubit gate scenarios can reach $O(2^{MN})$ \cite{samotij2015counting,jou2000number} (see \Cref{fig parallel sit}(d)). The crosstalk between qubits varies with each unique parallel gate scenario, and considering all possible scenarios results in an overwhelming number of frequency constraints. Given the exponential number of parallel scenarios, performing frequency configuration for all of them is infeasible. Thus, we focus on one two-qubit gate grouping pattern (see \Cref{fig parallel sit}(c)), dividing the two-qubit gates into four groups. Gates within each group can operate in parallel, but groups cannot execute in parallel with one another. This approach covers all possible two-qubit gates on the chip.

\begin{figure}[htbp]
\centering
\begin{minipage}[t]{0.2\textwidth}
\centering
\includegraphics[width=\textwidth]{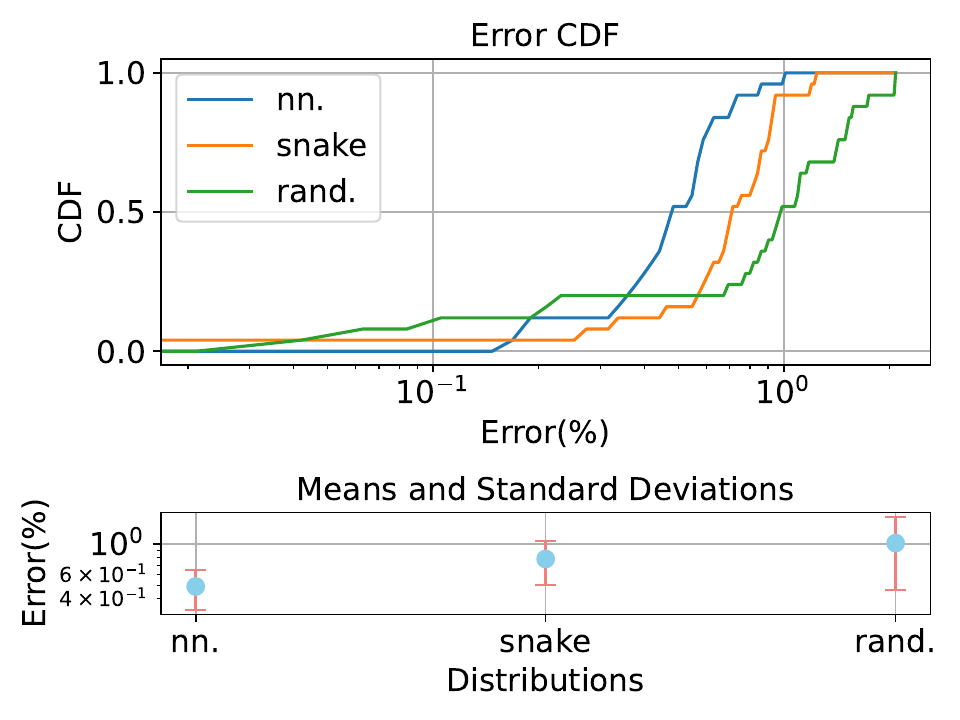}
(a)
\end{minipage}
\begin{minipage}[t]{0.2\textwidth}
\centering
\includegraphics[width=\textwidth]{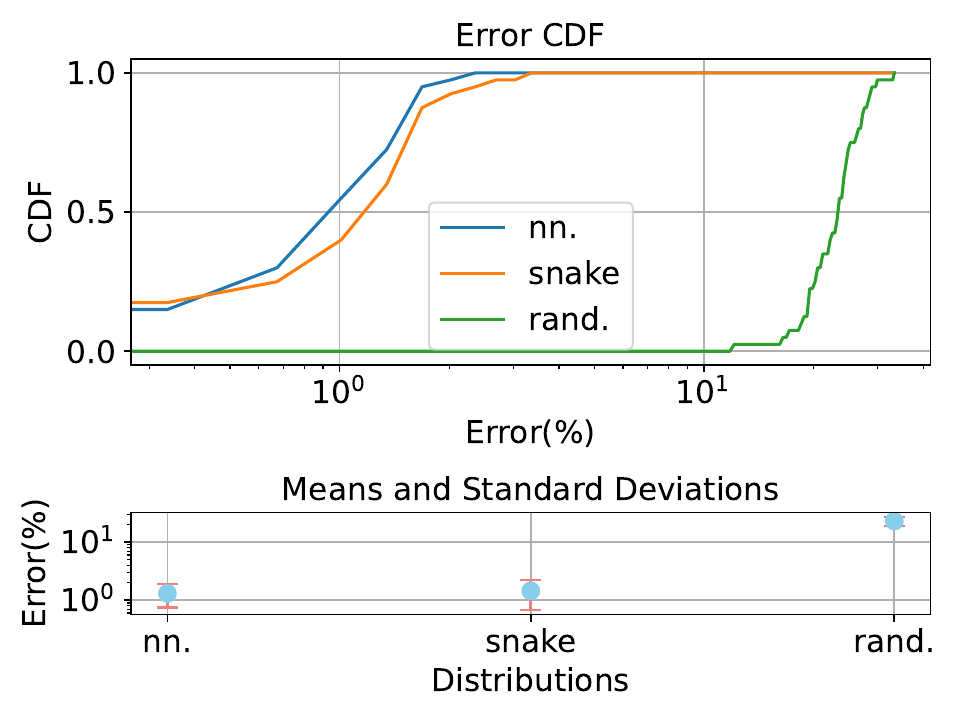}
(b)
\end{minipage}
\caption{
(a) The orange lines show the CDF of single-qubit gate errors after neural network-based optimization, the red line represents Google’s optimization, and the green line corresponds to a random configuration. The average errors are $0.49\%$, $0.78\%$, and $1.02\%$, respectively.
(b) For two-qubit gates, the orange lines show the CDF after neural network-based optimization, the red line represents Google’s baseline, and the green line corresponds to a random configuration. The average errors are $1.31\%$, $1.45\%$, and $23.3\%$, respectively.
}\label{fig rb xeb}
\end{figure}

\begin{figure*}[htbp]
\centering
\begin{minipage}[t]{0.3\textwidth}
\centering
\includegraphics[width=\textwidth]{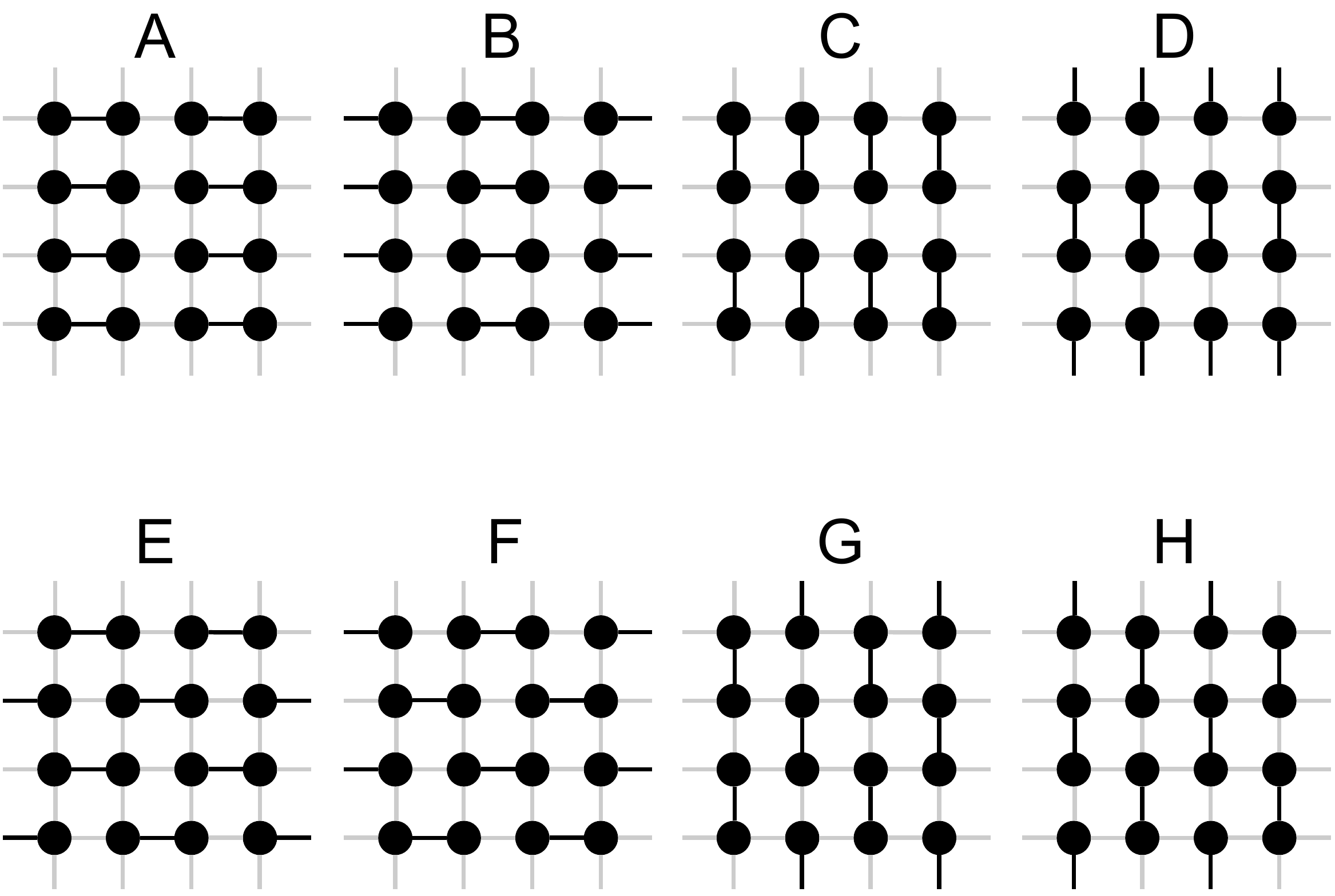}
(a)
\end{minipage}
\begin{minipage}[t]{0.18\textwidth}
\centering
\includegraphics[width=\textwidth]{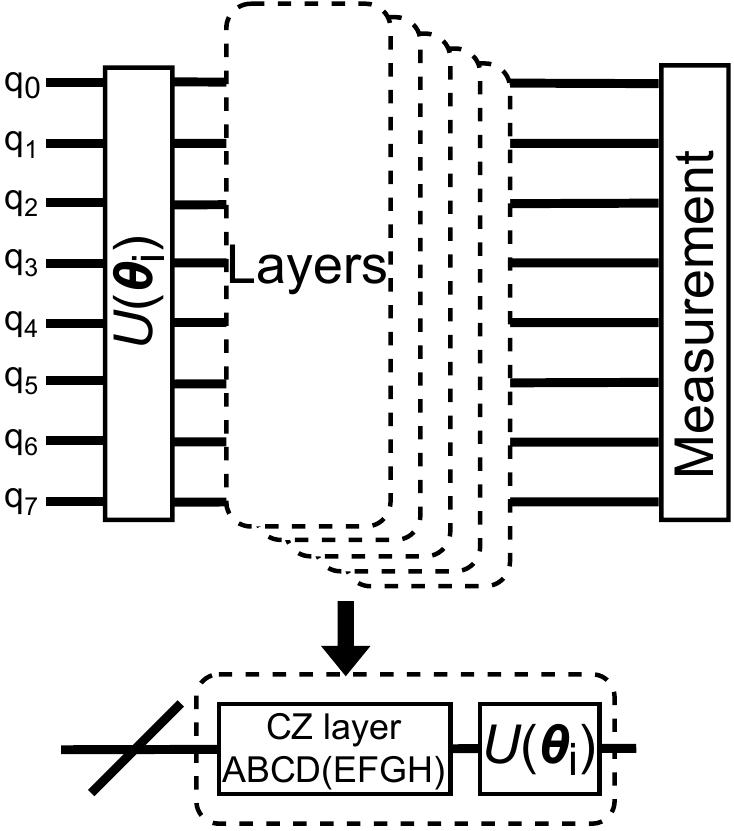}
(b)
\end{minipage}
\begin{minipage}[t]{0.3\textwidth}
\centering
\includegraphics[width=\textwidth]{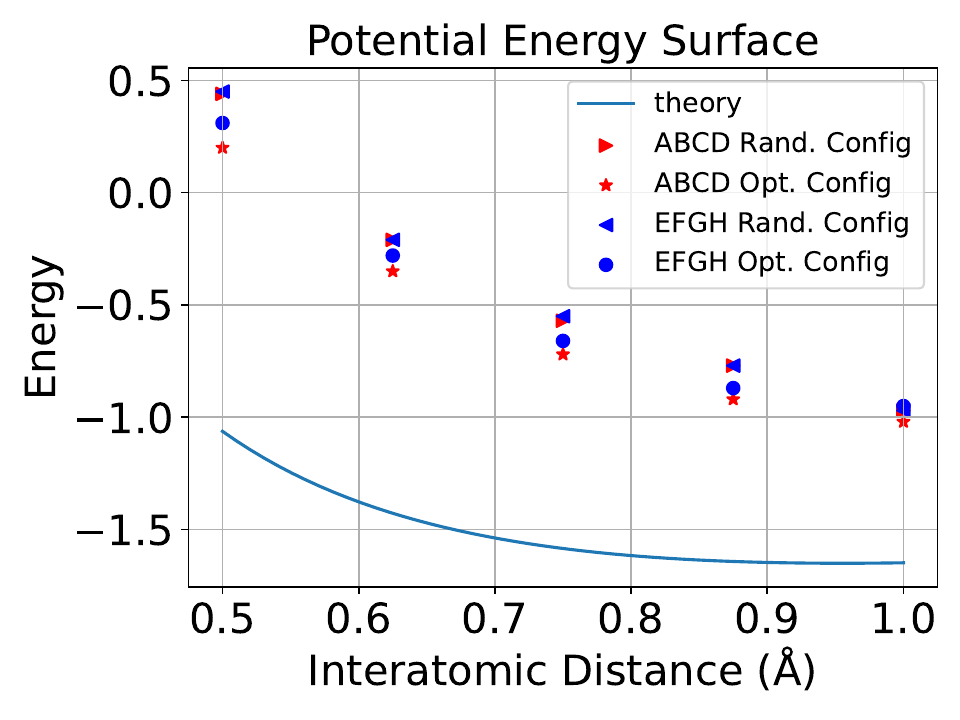}
(c)
\end{minipage}
\caption{
(a) Two sets of the largest parallelizable CZ gate patterns. 
Multiple CZ gates within any single ABCD (or EFGH) pattern can execute in parallel, while gates from different patterns cannot execute simultaneously.
(b) In the HEA circuit, each layer comprises trainable CZ and single-qubit gates arranged in the ABCD(EFGH) patterns, coupling all qubits together.
(c) The ground-state potential energy surface of the $\text{H}_4$ molecule was calculated using VQE circuits with the ABCD and EFGH ansatz patterns.
}\label{fig pat ans}
\end{figure*}

Using a trained neural network, we can estimate the error rates of all quantum gates based on their frequency configurations. Building on this, we propose an optimization scheme designed to identify the configuration that minimizes the error predicted by the neural network.

Each quantum gate operates at a designated frequency, and for an $M \times N$ chip, the number of frequency variables is $3MN - M - N$. 
The control system allows for a frequency precision of $\delta f$, so each frequency can vary within the range 
$F=\{f_{\min} + k \delta f |k \in \mathbb{N}, k \leqslant (f_{\max} - f_{\min})/\delta f\}$.  
This makes the problem scale exponentially as $O(|F|^{3MN})$, limiting us to incrementally optimize the frequency configuration within a local window $S$ on the chip.

Starting with a random configuration, we calculate the error for all quantum gates under this initial setup. Then, using a maximum optimization window size $S$, we iterate through all possible windows. 
For each window, we calculate the average gate error by summing the errors of all gates within the window and then optimize the frequencies of the two-qubit and single-qubit gates in the window with the highest average error. This iterative process continues until a configuration with minimal global average gate errors is achieved.

\Cref{fig freq err} illustrates the iterative optimization process for frequency configuration. \Cref{fig freq err}(a) shows the frequency configurations at the 1st, 2nd, and final (40th) iterations. In \Cref{fig freq err}(b), the predicted errors for each configuration are displayed. For each configuration, we calculate the average error of all quantum gates within each window $S$ of radius 2. The window $S$ with the highest average error is outlined in red and targeted for local optimization in the next iteration.

As seen in \Cref{fig freq err}(c-d), the average gate error across the chip gradually converges to below $10^{-2}$. 
Finally, by adjusting the radius of $S$, we observe that larger $S$ values result in lower converged average errors. 
However, as $S$ increases, the optimization process approaches global optimization, leading to a significant rise in computational complexity (see \Cref{fig freq err}(d)). 
Selecting an optimal $S$ allows the gate error to converge to a low level within a feasible computation time.
  
\Cref{fig rb xeb} shows the cumulative distribution function (CDF) of error probabilities measured after applying frequency configuration optimization. Single-qubit gates were evaluated using RB, while two-qubit gates were tested with XEB. Baselines include Google’s optimizer-based frequency allocation framework and a parallel experiment using random configurations. 
From \Cref{fig rb xeb}, our approach reduces average errors by $37.1\%$ and $52.0\%$ in the RB experiment, and by $9.7\%$ and $94.4\%$ in the XEB experiment, compared to Google’s method and random configuration, respectively, highlighting significant two-qubit gate optimization.

\textbf{Frequency configuration for HEA} The HEA is the most commonly used ansatz in VQE. 
The structure of this ansatz consists of a layer of single-qubit gates followed by a layer of entangling unitary operations $U_{ENT}$ that entangle all of the qubits in the circuit \Cref{fig pat ans}(b). 
Typically, these $U_{ENT}$ are composed of CNOT gates. Each time, a maximum set of parallelizable two-qubit gates is selected. 

Therefore, we need to consider the frequency configuration when designing the HEA. 
For each layer of parallel CNOT gates, the maximum allowable parallel set permitted by the configuration must be selected.

We optimized the frequency configuration of two-qubit gates based on the patterns ABCD or EFGH. The HEA ansatz follows a cyclic, multi-layer sequence: $\text{A}(\text{E})\rightarrow U(\bm{\theta_i})\rightarrow\text{B}(\text{F})\rightarrow U(\bm{\theta_i})\rightarrow\text{C}(\text{G})\rightarrow U(\bm{\theta_i})\rightarrow\text{D}(\text{H})$, where $U(\bm{\theta}_i)$ represents the parameterized single-qubit gate layer.

To verify the effectiveness of the frequency configuration optimization, we optimized the frequencies for two different patterns and compared the results with random frequency configurations. We also assessed the difference in the minimum energy of the ground-state potential energy surface (PES) from the theoretical line after frequency optimization. From \Cref{fig pat ans}(c), we conclude that, under the same ansatz, the potential energy surface for the random frequency configuration is higher than that of the optimized frequency configuration.
This demonstrates that the HEA designed with frequency-optimized configuration allows the quantum chip to operate at an optimal frequency, 
mitigating the impact of crosstalk on algorithm fidelity.

\textbf{Conclusion}
In this work, we developed a comprehensive strategy to optimize quantum gate frequencies on a multi-qubit chip, addressing a critical challenge in enhancing quantum computing performance. Combining neural network-guided prediction, frequency configuration, and rigorous benchmarking effectively reduced gate errors and minimized crosstalk in quantum circuits.

Our frequency configuration method, optimized for a specific two-qubit activation pattern, enables parallelizable two-qubit gates to execute with significantly lower error rates compared to unoptimized configurations. Experimental validation through Randomized Benchmarking (RB) and Cross-Entropy Benchmarking (XEB) demonstrated that our optimized configurations achieve the lowest error rates, validating the effectiveness of our approach. In testing with the Variational Quantum Eigensolver (VQE), circuits using the optimized pattern produced more accurate ground-state energy calculations for the $\text{H}_4$ molecular system. Notably, circuits with non-optimized configurations exhibited higher error rates, underscoring the importance of carefully designed frequency allocations.

Our findings highlight the potential of targeted frequency configuration and error minimization strategies in quantum computing, demonstrating how these methods can enable reliable performance improvements while managing computational complexity. Future research could extend this approach to larger quantum systems and explore dynamic frequency adjustments for real-time applications, further advancing the fidelity and scalability of quantum algorithms.

\textbf{Acknowledgements} 
This work was supported by the National Key Research and Development Program of China (Grant No. 2023YFB4502500).

\bibliography{apssamp}

\end{document}